# Manuscript
*The impact of Electricity Blackouts and poor infrastructure on the livelihood of residents and the local economy of City of Johannesburg, South Africa.*

Authors: Nkosingizwile Mazwi Mchunu[1;] George O Onatu[2] & Trynos Gumbo[3]
Faculty of Engineering and Built Environment, Department of Urban & Regional Planning, University of Johannesburg, South Africa,

mailto:mazwi.s65@gmail.com[1]; gonatu@uj.ac.za[2] & tgumbo@uj.ac.za[3]

**Abstract:**
This paper discusses the impact of electricity blackouts and poor infrastructure on the livelihood of residents and the local economy of Johannesburg, South Africa. The importance of a stable electricity grid plays a vital role in the effective functioning of urban infrastructure and the economy. The importance of electricity in the present-day South Africa has not been emphasized enough to be prioritized at all levels of government, especially at the local level, as it is where all socio-economic activities take place. The new South Africa needs to redefine the importance of electricity by ensuring that it is accessible, affordable, and produced sustainably, and most of all, by ensuring that the energy transition initiatives to green energy take place in a planned manner without causing harm to the economy, which might deepen the plight of South Africans. Currently, the City of Johannesburg is a growing spatial entity in both demographic and urbanization terms, and growing urban spaces require a stable supply of electricity for the proper functioning of urban systems and the growth of the local economy. The growth of the city brings about a massive demand for electricity that outstrips the current supply of electricity available on the local grid. The imbalance in the current supply and growing demand for electricity result in energy blackouts in the city, which have ripple effects on the economy and livelihoods of the people of Johannesburg. This paper examines the impact of electricity blackouts and poor infrastructure on the livelihood of residents and the local economy of Johannesburg, South Africa.

## 1. Introduction

Energy has been a major factor throughout human civilization as it has been used to make life bearable for the human species (Osipov et al., 2014). On the timeline of human development, the 18th century represented a great leap forward as the result of the invention of the steam engine, which brought about the first industrial revolution (De Pleijt et al., 2018). The steam engine used coal as the source of energy (Wrigley, 2013). which brought about many changes in the way the world works from traveling on land and sea as the travel time became shorter and more reliable. That created a ripple effect on the creation of urban settlements and cities due to the revolutionary effect of energy. This saw the transformation of agrarian communities into urban settlements with diverse economic activities (Li et al., 2019) . It is for this reason that we believe energy is fundamental to urban planning, and access to it is significant to the failure or success of the city. For the purposes of the current scope of this research, we shall be limited to focusing on the city of Johannesburg. The City of Johannesburg has evolved from what was a small mining town over a century ago into an economic hub of the African continent (Abrahams and Everatt, 2019). The 21st century has brought so many changes in terms of development and technological innovations (Smiths, 2002), and energy is at the center of all innovation. These changes are mainly driven by the interconnected globalized world economy of cities, where global cities collaborate with each other in terms of skills, services, and products that are being produced (Ambos et al., 2021). The City of Johannesburg is part of this global community of cities that are driven by not only collaboration but also ruthless competition as one place leverages the advantages it has over the other. That competition spills over to these cities to prosper and attract investments, and some cities suffer decline and are trapped in the perpetual cycle of high unemployment, poverty, and inequality. The center of the research will be focusing on electrical energy, as the word energy has diverse forms and meanings. Without energy, cities do not function properly. It has been observed in the power outage that started back in 2008 that what is more noteworthy is the disruption that is caused by limited access to electrical energy (Kaseke and Hosking, 2013).

## 2. Background

The City of Johannesburg has evolved from what was a segregated city across racial lines into a non-racial urban space politically, socially, and economically (Van Eden, 2015).These changes have brought about new dynamics under democratic dispensation to recreate an inclusive city. The City of Johannesburg has attempted to reshape the existing and emerging spatial patterns of a divided, sprawling city to recreate a space of equal access to opportunities for everyone who gets to interact with the city (Todes, 2012). The reshaping of the city requires spatial planning and infrastructure development initiatives, which will enable the growth and expansion of the city's economy and improve the lives of people that reside in the city. The improvement of the quality of life means that the access to basic services (Nnazodie, 2013) , by communities residing in the different parts of the city should be the same regardless of class or economic status, and the delivery standard of public services should be the same for every city dweller in the City of Johannesburg. Presently, the city grappled with problems that were caused by discrimination, segregation, and racial inequality and manifested themselves in poor and substandard services for the majority of the city dwellers. Presently, the city is facing new and emerging challenges as it continues to grow and requires to meet the electricity demand of its growing population (Wright and Calitz, 2020). The focus of the study is on the impact of access to energy on the lives and economy of the city of Johannesburg. The city of Johannesburg is the largest economy in the country (Abrahams and Everatt, 2019), as it accounts for a large percentage of the GDP of South Africa. According to Statistics South Africa," Johannesburg's status as an engine of the national economy contributes to 16% of South Africa's GDP". The status of the city of Johannesburg as a leading economic contributor to the South African GDP requires energy to sustain it, as it has been highlighted earlier that energy is the cornerstone of all civilizations and human progress throughout history (Caineng, 2022) , and that this continues to the present day. This research study will focus on energy issues in relation to access, with the objective of determining the impact caused by access on the lives and economy of the City of Johannesburg. At a time when there is much debate about the need to diversify energy sources and move away from fossil fuels, it is important for developing nations to investigate basic issues relating to access to energy and its role in society. The figure below shows the maps of the city of Johannesburg, Source: (Chirisa and Matamanda, 2019)

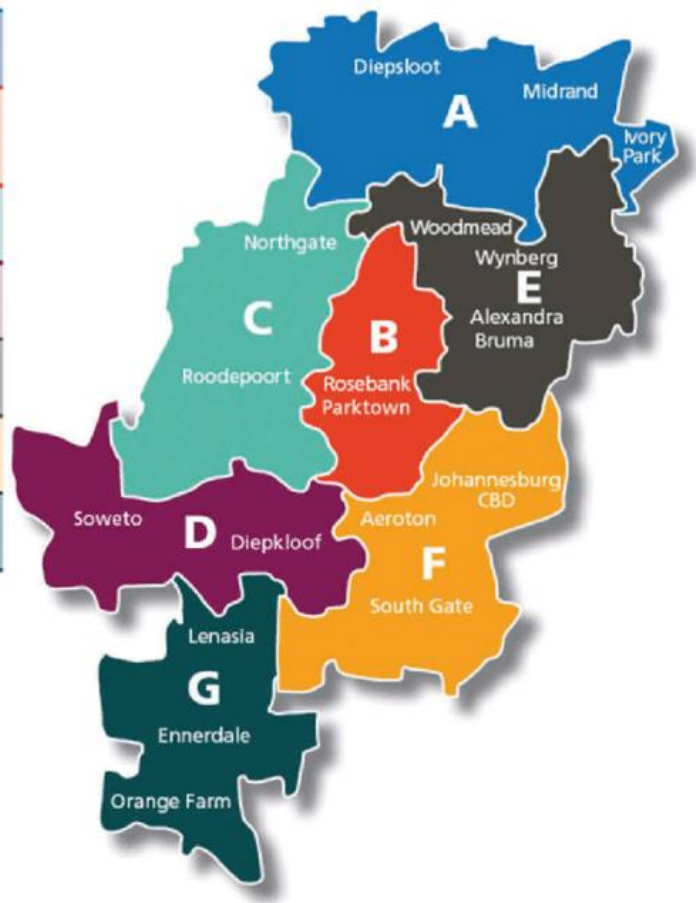

## 3. Theoretical Frameworks

The analogy of a structure or a building with a foundation applies in the context of the research as well. The theoretical framework is the foundation from which all knowledge is constructed (metaphorically and literally) for a research study (Osanloo and Grant, 2016). In other words, the theoretical framework serves the purpose of providing structural support for the study. The absence of a clearly defined theoretical framework weakens the purpose of the study as it leaves it without a firm footing to provide a clear rationale. From a widely held belief in academia that all forms of research are theoretical in nature, that fundamental belief on its own places a strong significance on the theoretical framework as the anchor from which ideas emanate (De Hass, 2021). Thus, the theoretical framework consists of the selected theory (or theories) that undergirds thinking around the study concerning how the topic is understood and how you plan to research it, as well as the concepts and definitions from that theory that are relevant to your topic. In the context of this study, there are three theories that serve the purpose of the theoretical framework, namely: (a) right to the city; (b) collaborative planning theory and (c) production of space. The next section will look at the theoretical frameworks that guide this research.

### 3.1 Right to the city

The right to the city manifests itself as a superior form of rights: the right to freedom, to individualization in socialization, to habitat, and to inhabit (King, 2019). Taking into consideration the historical legacy of segregation in South Africa, where the equal right to public spaces in urban areas was significantly infringed upon by laws that were in place up until 1994, in 1967, Henri Lefebvre described the right to the city as a "cry and demand." In the context of South African urban spaces, the right to the city represents the need of the majority of the population that was previously marginalized to be freed and to be able to access the benefits of the city in terms of quality infrastructure and the opportunities that come with it. Almost three decades later, after the new democratic dispensation has emerged, the spatial patterns of the city continue to reflect the legacy of segregation. It is for this reason that the right to the city is argued not fully realized by the people of the City of Johannesburg, including access to electricity.

### 3.2 Production of Space

The theory that undergirds the rationale of the study emanates from Herni Lefebvre's seminal work called "Theory of the Production of Space," whereby the production of space has three dimensions, namely: (a) spatial practice, (b) representations of space, and (c) representational space. Each dimension of the production of space has three categories: (a) subjects, (b) objects, and (c) activities (Fuch, 2018). The production of space has elements that make up the fabric of society; consequently, it follows the social system approach from a functional perspective. The study seeks to investigate the impact of a lack of access to energy (electricity) on the subjects, objects, and activities within a social space of the local municipality called the City of Johannesburg Metropolitan Municipality.

### 3.3 The Theory of Collaborative Planning

Collaborative planning theory is a theory rooted in democracy and participation and influenced by four traditions: social reform, policy analysis, social learning, and social mobilization (Beauregard, 2020). It aims to promote positive reforms and optimal solutions in planning ideas and best practices, involving community members at the grass-roots level. This form of social leaning involves community members being represented in developmental decisions, allowing them to be informed about authorities' plans and visions. Collaborative planning also fosters social mobilization, allowing communities to unite and take decisions on planning (Horn, 2021). It is a democratization of planning, allowing diverse thoughts and visions to be included in urban development in the 21st century (Fletcher, 2020). Collaborative planning theory brings together the state, private sector, civil society organizations, and the public to bring about collaboration in planning space and inclusion in state actions. It decentralizes power away from the state and key stakeholders, allowing community members to have input in planning stages and decisions that affect their lives. This bottom-up approach ensures the inclusion of all relevant people affected by planning decisions.

### 4. Methodology

The systematic nature of research requires a scientific methodology to be valid (Pigott, and Polanin, 2020), whereby a certain methodology has to be followed when collecting data to substantiate the research. Research methodology simply refers to the practical aspect in terms of how any given portion of research or investigation is conducted (Kothari, 2004). More specifically, it is about how a researcher systematically designs a study to ensure valid and reliable results that address the research's aims and objectives.

### 5. Research Approach

The research is a mixed-methods approach, it combines the use of qualitative and quantitative methods in a single study namely: Qualitative approach and Quantitative approach.

### 5.1 The qualitative approach

It is an approach based on getting an in-depth understanding from the population's perspective in a descriptive form (Mahajan, 2018). It involves stories from people talking about how the problem affects them as individuals, and it is not only limited to stories; it also includes observations, online surveys and focus groups discussion.

### 5.2 Quantitative approach

This is an approach that emphasizes objective numerical measurements and the statistical analysis or numerical analysis of data collected through polls, questionnaires, and surveys, or by manipulating pre-existing statistical data using computational techniques (Satinem, 2016) .Applicable to this investigation could be statistical data showing the current price of electricity, the quantity of electricity that is available to serve the City of Johannesburg both in the past and present.Therefore research will follow a mixed method approach to collect both qualitative data and quantitative data.

## 6. Sampling Method

A 'sample' is a subset of the population selected to be representative of the larger population. Since the research cannot study the entire population, we need to take a sample. The design of the sample will be within the sampling area where the investigation will take place, the City of Johannesburg metropolitan municipality in Gauteng province, South Africa. The research uses both probability sampling and non-probability sampling. Whereby the business community and ordinary citizens of the city will be key participants in the study in terms of providing input when it comes to how they are impacted by the lack of access to electricity. According to Statistics SA (2021), the current metro area population of Johannesburg in 2021 will be 5,927,000, a 2.33% increase from 2021 (Marias et al., 2022). To ensure the rigorous procedure of sampling is followed, there are two main types of sampling procedures that will be followed, namely: (a) probability sampling and (b) nonprobability sampling, as the result the online surveys were conducted with key informants and not participants direct observations. There were 30 participants that took part in in the study through purposive convenience sampling.

## 7. Data Collection

Data collection is the process of gathering and measuring information about variables of interest in research (HR & Aithal, 2022), whereby the collection takes place in an established systematic fashion that enables the researcher to arrive at findings and be in a position to answer stated research questions. Secondly, data collection enables the research to test hypotheses and evaluate overall outcomes (Belcher et al., 2020). Data collection is a key component of research in all fields of study (Singh & Thurman, 2019). While methods vary by discipline, the emphasis on ensuring accurate and honest collection remains the same. The research requires first-hand information, including the one that the researcher collects, which is regarded as primary data, and information that has been collected from other sources and published, which is regarded as secondary data collection methods. Research participants primary data was conducted and collected based on order restrictions of COVID 19 regulations with restrictions on face-to-face contact, as the result the online surveys were conducted with key informants and nonparticipant direct observations. Secondary data was collected through reviewing literature from the internet, textbooks, journals, and government publications. Primary data types, namely:

1. Online Surveys
2. Observations

This is the type of data that is collected specifically for a particular research problem.

## 8. Results

Quantitative analysis for the study is a descriptive analysis, as quantitative data is expressed in numerical form. The purpose of quantitative data analysis is to make a deduction out of a survey with a goal of answering the research question.

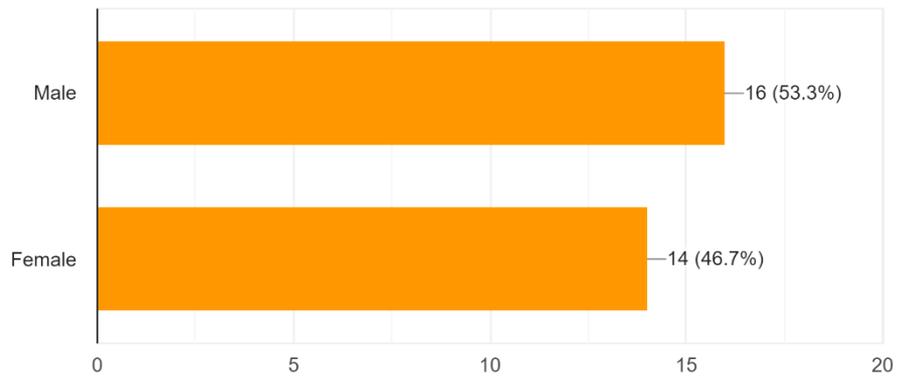

SECTION A BACKGROUND AND INFORMATION This section of background information is about finding more about the respondent, and we promise to keep it anonymous.
30 responses

- Gender differences in participation: 53% male and 47% female (see figure 8.1)

3. Ethnicity
30 responses

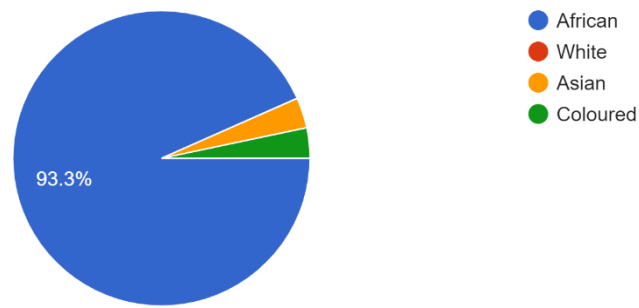

- Population demographics: 93% Africans, 3% coloured, and 3% Asians (see figure 8.2 Above)

SECTION B: THE FACTORS THAT CAUSE LACK OF ACCESS TO ELECTRICITY TO PEOPLE LIVING IN THE CITY OF JOHANNESBURG 7. What are reported co...ricity black) outs in your area where you live?
30 responses

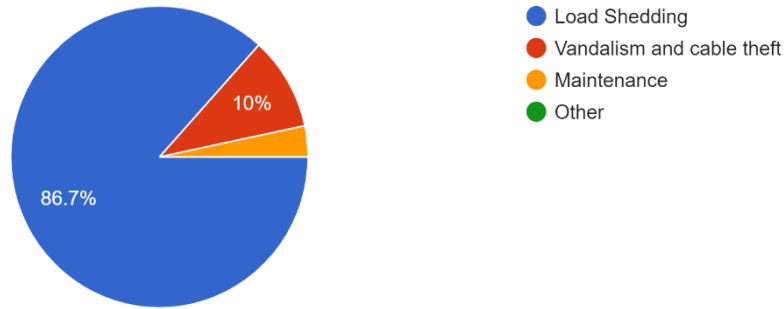

- Factors that cause lack of electricity: 87% load shedding in residential areas and 90% in industrial areas (see figure 8.3 above).

14. How does energy blackouts affect your domestic life and personal life?
30 responses

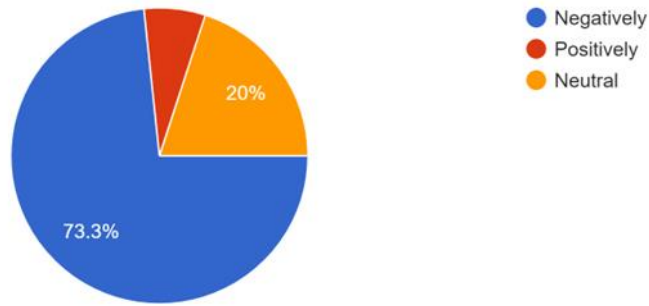

- Impact of a lack of electricity on your domestic life and work place: 73% negative impact on domestic and 97% negative impact at your place of work (see 8.4). For domestic life includes. Cooking, washing, ironing, watching the television, lighting, reading and entertainment. Whereas for work place includes, meetings, conferences, typing report on the computer, lighting, preparing coffee, presentations, and internet services and intercom connections. This result also supported by recent report by Sunday Times of November 12, Business Section that Power cut or loadshedding pose significant challenges to small businesses-which account for a third of the country's GDP.

13. How does energy blackouts affect our line of work in terms of productivity and your ability to preform your work ?
30 responses

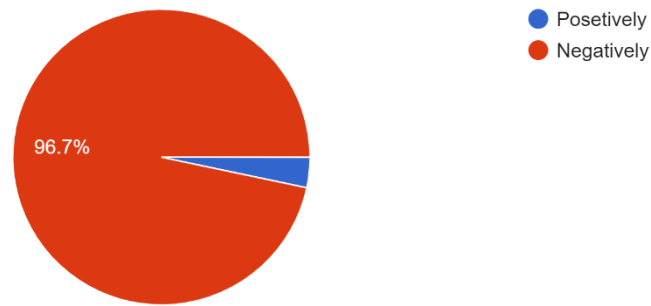

- Impact of a lack of electricity on work 97% negative impact at your place of work (see 8.5)

- The data indicates a strong relationship between loss of productivity and lack of electricity, whereby during an energy blackout there is a massive loss of productivity and inconvenience that comes with it in the form of disruption in domestic and communication services. This result is also supported by the work of statistics South Africa over the course over the years.

### 9. Conclusions

Conclusion on the research findings: This section will draw the major conclusion of each research objective that was stated at the beginning of the study, which are namely

**9.1 To assess the impact of lack of access to electricity on the lives of people and the economy of the city of Johannesburg**

The impact due to lack of access to electricity is negative, based on the primary data collected, whereby over 90% of participants indicated the ways in which blackouts affect their lives in a negative way and also contribute to the loss of productivity.

**9.2 To examine the economic effects of reliable and accessible electricity on the lives of people and the economy of the City of Johannesburg.**

Based on the negative impact socially, as indicated in 5.2.1, the impact is also negative economically. On the other hand, when electricity is accessible and reliable, people are able to go about their daily lives and be productive by pursuing their different endeavours in different sectors of the economy.

**9.3 To evaluate the socioeconomic effects of electricity on the lives and people of the City of Johannesburg.**

The effect of a loss of productivity contributes to the overall loss of income that would have been generated and circulated within the economy. The sum of it is the part of the revenue that the city was in a position to collect through the sale and consumption of electricity. As a result the

City's loss before tax grew up to 21.87% during 2020/ 2021 financial year compared to previous financial year of 2019/2020 (CPAIR, 2021).

**9.4 To assess strategies of the Department of Energy that are aimed at improving the generation and adequate supply of electricity.**

At the national government level, there have been numerous initiatives that are state-led and geared towards empowering municipalities to generate their own electricity and have a self-sufficient energy grid. The amendment of the Electricity Act to allows municipalities to generate electricity as independent entities within the framework of the law. It is one of the key strategies that seeks to empower municipalities to secure electricity supply.

**9.5 To assess the challenges faced by energy generation development and access in the city of Johannesburg.**

One of the challenges is with regard to the mobilization of resources by the City of Johannesburg to move towards harnessing the potential it has for generating electricity. The City is struggling financially and this is noted widely by the former Mayor Mpho Phalatse who noted.that the " City finances are not in best position, but council taking steps to rectify it" (News, 24 of 2022). The second challenge is the urban planning practice that seems to underestimate the potential of the municipal authority when it comes to passing and enforcing laws that will move communities to adopt a sustainable energy mix by allowing residential areas to install solar panels that are connected to the local electricity grid to keep it stable.

## 10. Recommendations

Based upon the study conducted, this section outlines the recommendations that can be adopted by the City of Johannesburg in terms of energy policy and a holistic approach to energy as a whole, with the objective of insulating itself from the risks of experiencing disruptive and counterproductive power cuts. By following the existing legal framework and other legislative mechanisms that are available to enable local governments to develop.

**The city must prioritize the importance of having a stable economic environment.**

The city must seek to prioritize having a stable electric grid, which is necessary for a well-functioning economy. To ensure that the city continues to be a competitive environment for the economy that will create employment and contribute to the overall tax base that the city has the potential to collect. The prioritization of having a stable electric grid within the local municipality means that the city must acknowledge its limited resources and actually form partnerships with other private organizations for mutual benefit that will be useful to both citizens and the local economy.

**Using the neighbourhood to generate renewable energy to improve access to affordable electricity.**

The city could use the legislative powers it has by passing bylaws that will enable every household to have solar panels on their roofs in order to maximize and improve the amount of electricity that is available on the grid. The strategy will ensure that households that cannot pay for electricity can take an active role in at least producing electricity through their roofs and hoses. To this end the existing City of Johannesburg Metropolitan Municipality By-laws of 2016 need to be reviewed to accommodate this reality. Improve electricity infrastructure investment at the local government level. The improvements to the electricity infrastructure investments will enable the city to be in a position to carry out maintenance work on the local electricity grid. The investment includes the recruitment of artisans from different skill sets that will be dedicated to supporting electricity infrastructure, followed by engineers from different disciplines for planning and monitoring purposes.